\documentstyle[aas2pp4]{article}

\slugcomment{Submitted to ApJ, Part I}

\lefthead{Gibson}
\righthead{Nucleosynthesis \& the IMF Upper Mass Limit}

\begin{document}

\title{Can Stellar Yields Accurately Constrain the Upper Limit to the
Initial Mass Function?}

\author{Brad K. Gibson}
\affil{Mount Stromlo \& Siding Spring Observatories, 
Australian National University, Weston Creek P.O., Weston, ACT, 
Australia  2611}

\def\spose#1{\hbox to 0pt{#1\hss}}
\def\simlt{\mathrel{\spose{\lower 3pt\hbox{$\mathchar"218$}}
     \raise 2.0pt\hbox{$\mathchar"13C$}}}
\def\simgt{\mathrel{\spose{\lower 3pt\hbox{$\mathchar"218$}}
     \raise 2.0pt\hbox{$\mathchar"13E$}}}
\def\eg{{\rm e.g. }}
\def\ie{{\rm i.e. }}
\def\etal{{\rm et~al. }}

\begin{abstract}
Recent determinations of the upper mass limit to the 
local initial mass function (IMF) claim a value of $m_{\rm U}=50\pm 10$
M$_\odot$, based upon direct comparisons of the observed oxygen and 
iron abundances
in metal-poor stars with the predicted stellar yields from Type II supernovae 
(SNe).
An unappreciated uncertainty in these analyses is the input physics intrinsic
to each SNe grid, and its effect upon stellar nucleosynthesis.
We demonstrate how such uncertainties, coupled with the uncertain
metal-poor halo star normalization, while allowing us to set a \it lower \rm
bound to $m_{\rm
U}$ of $\sim 40$ M$_\odot$, nullifies any attempt at
constraining the \it upper \rm bound.
\end{abstract}


\section{Introduction}
\label{introduction}

The initial mass function (IMF), a measure of the distribution of masses,
at formation, of a given stellar generation, is one of the key components of
galaxy evolution modeling.  
Despite its importance, accurate determination of the IMF
remains one of the most elusive problems in modern astronomy.  This 
elusiveness is manifested both in ongoing attempts to understand
its physical underpinings (\eg Padoan, Nordlund \& Jones 1997),
as well as simply
characterizing its shape and mass limits from an observational tack (\eg 
Kroupa, Tout \& Gilmore 1993; Scalo 1986).

In its simplest form, the IMF can be considered a power law of the form
$m\phi(m){\rm d}m\propto m^{-(1+x)}{\rm d}m$, where $m\phi(m){\rm d}m$ is the
number of stars born in the mass interval $m\rightarrow m+{\rm d}m$.  The goal
for theorists and observers, alike, is then the determination of the slope of
this function $x$, as well as its upper and lower limits ($m_{\rm U}$ and
$m_\ell$, respectively).

While the slope of the IMF, at least in the solar neighborhood (and for
masses greater than a few solar masses), would appear to
lie somewhere in the range $x\approx 1.3$ (Salpeter 1955) to $x\approx 1.7$
(Kroupa, Tout \& Gilmore 1993), and the lower mass limit is close to
$m_\ell\approx 0.2$ M$_\odot$ (Bahcall \etal 1994), the upper mass limit $m_{\rm
U}$ still remains \it highly \rm uncertain.  Even a cursory examination
of the literature corroborates this point, with
values in the range $m_{\rm U}\approx 20\rightarrow 200$ M$_\odot$ suggested
by a variety of direct and indirect techniques (\eg Maeder \& Meynet 1989;
Klapp \& Corona-Galinda 1990; Pagel \etal 1992; Maeder 1992; Massey \etal 
1995; Kudritzki 1997).

A hybrid approach to determining $m_{\rm U}$,
combining predictions
of the \it theoretical \rm yields from Type II supernovae (SNe), with
the \it observed \rm abundances in the old metal-poor stars (\ie those
which bear the clear
imprint of yield ``pollution'' from these same SNe, with no ``dilution'' from
Type Ia SNe, whose progenitor lifetimes are considerably longer than the Type
II timescales), has been 
the subject a recent series of
papers (Tsujimoto \etal 1995,1997; Yoshii, Tsujimoto \& Nomoto 1996).
The premise here is that because Type II SNe
$\alpha$-element (\eg O, Mg, Ne) yields are a strong function of progenitor
mass, whereas products of explosive burning (\eg Fe, Si, Ca) are less so, the
IMF-weighted average of their ratios must necessarily 
also depend strongly upon $x$ and $m_{\rm U}$.  Tying these yield ``averages''
to the halo abundances then, 
in principle, provides a unique indirect probe on the upper
mass limit to the IMF.\footnote{Tsujimoto \etal (1997)
extend this ``hybrid'' approach to simultaneously constrain the IMF
mass limits $m_{\rm U}$ and $m_\ell$, as well as the slope $x$.  We shall only
be concerned with the $m_{\rm U}$ determination in what follows, primarily for
brevity, but also because the lower
mass constraint rest squarely upon uncertain photometric
calibrations.}

Following this technique, Tsujimoto \etal (1997) recently concluded
that the upper mass limit to the IMF in the solar neighborhood is $m_{\rm
U}=50\pm 10$ M$_\odot$.  
To do so, they made explicit use of the Tsujimoto \etal (1995)
compilation of Type II SNe yields.
What was not fully appreciated in their study though
was just how dependent their result was to this
\it particular \rm yield compilation and the adopted halo abundance
normalization.
It is to this lack of appreciation that our current study is
addressed.

After providing a minimal introduction to the model ingredients in Section
\ref{analysis1}, we demonstrate in Sections \ref{analysis2} and \ref{analysis3}
that this technique results in
$m_{\rm U}=50\pm 10$ M$_\odot$ \it only \rm for the Tsujimoto \etal (1995)
yields, combined with a halo normalization of [O/Fe]$_{\rm h}=+0.41$.  
Duplicating the analysis with ``competing'' yield compilations which
sample a wide variety of convection and mass-loss treatments (\eg Woosley \&
Weaver 1995; Langer \& Henkel 1995; Arnett 1996), clearly demonstrates that
Tsujimoto \etal (1997) have significantly underestimated the uncertainty
associated with thir determination of $m_{\rm U}$.
Our results are summarized in Section \ref{summary}.

\section{Analysis}
\label{analysis}

\subsection{The Basic Formalism}
\label{analysis1}

The observed oxygen-to-iron abundance ratio in halo dwarfs and giants,
[O/Fe]$_{\rm h}$, can be linked (see Tsujimoto \etal 1997 -- hereafter, T97 --
for details)
to the theoretical Type II SNe yields $m_{\rm O}^{\rm ej}$ 
and $m_{\rm Fe}^{\rm ej}$ via
\begin{equation}
\biggl({{\rm O}\over{\rm Fe}}\biggr)_{\rm h} =
\biggl({{\rm O}\over{\rm Fe}}\biggr)_\odot\times 10^{\rm [O/Fe]_{\rm h}} =
{{\int_{10}^{m_{\rm U}}m_{\rm O}^{\rm ej}m^{-(1+x)}\,{\rm d}m}\over
{\int_{10}^{m_{\rm U}}m_{\rm Fe}^{\rm ej}m^{-(1+x)}\,{\rm d}m}},
\label{eq:imf1}
\end{equation}
\noindent
where the coefficient (O/Fe)$_\odot\equiv$7.55 is the solar (meteoritic)
mass fraction ratio from Anders \& Grevesse (1989), and $x$ is
the slope of the IMF (recall Section \ref{introduction}).

By specifying the low metallicity
``plateau'' value for [O/Fe]$_{\rm h}$ (\eg Figure 11 of
Timmes, Woosley \& Weaver 1995; Figure 2 of Bessell, Sutherland \& Ruan 1991), 
and the IMF slope
$x$, equation \ref{eq:imf1} 
provides a unique upper mass limit $m_{\rm U}$\it,
for a given Type II SNe yield compilation\rm.

\subsubsection{Ingredients}
\label{analysis1a}

Essentially, there are only two input
ingredients which interest us, in what follows.
First, the source of Type II SNe yields (right-hand side of equation
\ref{eq:imf1}), and second, the halo's ``plateau'' [O/Fe]$_{\rm h}$ (left-hand
side of equation \ref{eq:imf1}).
Let us briefly comment on the latter ingredient first.

There is no debate as to the reality of the $\alpha$-element
overabundance seen in halo dwarfs and giants (\ie [O/Fe]$_{\rm h}\simgt +0.2$), 
but there does remain a factor of $\sim 2$
uncertainty as to its asymptotic ``plateau'' value for [Fe/H]$\simlt -3$.
T97 adopt [O/Fe]$_{\rm h}=+0.41$\footnote{T97's claim to adopt [O/Fe]$_{\rm
h}=+0.41$ is 
not entirely correct, as their code does not follow O and Fe, per
se, but only the isotopes $^{16}$O and $^{56}$Fe, thereby underestimating the
true iron yield by $\sim 10$\% (and oxygen by a smaller
amount).  This has the effect that 
instead of using (O/Fe)$_\odot=7.55$ in equation \ref{eq:imf1}, T97
use (O/Fe)$_\odot=8.20$ (\ie ($^{16}$O/$^{56}$Fe)$_\odot=8.20$).  By itself,
this is not a problem; the problem arises in that T97 then normalize equation
\ref{eq:imf1} to the observed halo ratio [O/Fe]$_{\rm h}=+0.41$ - this observed
ratio, though,
reflects the \it total \rm O and Fe abundances, and \it not \rm just the
isotopes $^{16}$O and $^{56}$Fe.
In practice, what this means is that T97 use (O/Fe)$_{\rm h}=21.1$ (recall
equation \ref{eq:imf1}), which corresponds to [O/Fe]$_{\rm h}=+0.45$, and \it
not \rm +0.41.  To truly normalize their models to [O/Fe]$_{\rm h}=+0.41$, T97
should have used (O/Fe)$_{\rm h}=19.4$ \it and \rm included the total O and Fe
for the yields in the right-hand side of equation \ref{eq:imf1}.}
but it is apparent that
this depends somewhat upon the sample selection.  Visual inspection of Figure
11 of Timmes \etal (1995), Figure 2 of Bessell \etal (1991), and Figure 1 of 
T97, each drawn from slightly different sources,
shows that the plateau
lies roughly between [O/Fe]$\approx +0.4\rightarrow +0.6$; \ie T97's
value of +0.41 appears to lie on the lower end of the plateau 
``distribution''.  T97 did not
demonstrate how the predicted $m_{\rm U}$ changes as a function of this halo
``normalization'', a point to which we return in Section \ref{analysis3}, where
we duplicate their analysis using [O/Fe]$_{\rm h}=+0.60$ (\ie that 
favored by Bessell \etal 1991).

\begin{figure}[ht]
\epsscale{1.0}
\plotone{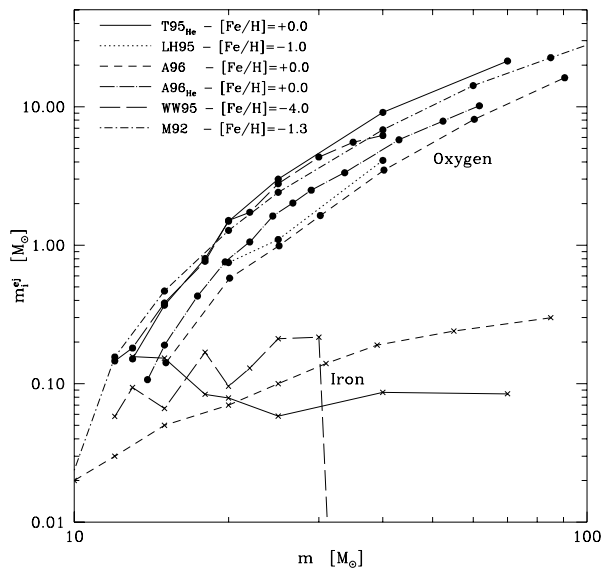}
\caption[fig1.eps]{
Oxygen (upper four curves) and iron (lower three curves)
yields as a function of progenitor mass, for the primary yield
compilations considered here -- T95=Tsujimoto \etal (1995); LH95=Langer \&
Henkel (1995); A96=Arnett (1996); WW95=Woosley \& Weaver (1995); M92=Maeder
(1992).  The models of M92, LH95 and A96
are only evolved up to the completion of oxygen burning, hence an iron yield is
not directly associated with either grid.  We indirectly associate Arnett's
(1991) iron yields with his newer 1996 oxygen yields.  See text for details.
\label{fig:fig1}}
\end{figure}

The Type II SNe yields are the primary input into equation \ref{eq:imf1}.
T97 adopt their earlier 1995 yield compilation (T95,
hereafter), which itself
is an offshoot of the Thielemann, Nomoto \& Hasimoto
(1996) models.  
The solid curves in Figure \ref{fig:fig1} represent the T95 oxygen and iron
yields adopted in the T97 analysis.

It is readily apparent though, as even a cursory glance at Figure 1 should
show, that there still remain \it substantial \rm
uncertainties in the yield computations for Type II SNe.  These differences
have already been described in detail by Arnett (1995), Langer (1997), and
Gibson, Loewenstein \& Mushotzky (1997), to name just a few.

T97 assume that these differences can be reconciled by considering
what they (mistakenly) assume is the maximal deviation from their canonical 
(T95) yields.  This deviation is taken to be a mass-independent 30\% excess
added to their oxygen yields; iron, though, is assumed to be invariant.
T97 justify this 30\% excess as the ``extreme'', since Woosley \& Weaver's
(1995) (WW95, hereafter) [Fe/H]=+0.0 oxygen yields are roughly 30\% greater
than those of T95 (also [Fe/H]=+0.0).\footnote{Actually, WW95's [Fe/H]=+0.0
oxygen yields (their favored Case B energetics)
are greater than those of T95 \it only \rm for $m\simlt 30$ M$_\odot$.}
Herein lies several (intertwined) problems.

First, solar metallicity yields are inappropriate for the discussion at hand;
T97 attempt to constrain $m_{\rm U}$ by comparing the abundance
ratios in metal-poor stars (\ie [Fe/H]$\simlt -2$) with the 
yields predicted from [Fe/H]=+0.0
Type II SNe models.  A more appropriate course 
of action is to consider a much lower
metallicity grid of models.  Because T95 do not compute [Fe/H]$<$+0.0 models,
T97 could not explicitly consider metallicity effects.

If we compare (properly)
T95's oxygen yields with WW95's sub-solar metallicity grid, the 30\%
``excess'' adopted by T97 is no longer relevant.
Figure \ref{fig:fig1} shows that WW95's [Fe/H]=-4.0 models
have oxygen yields virtually indistinguishable to the [Fe/H]=+0.0
models of T95, except for $m\simgt 35$ M$_\odot$, where WW95 now lies
systematically \it below \rm T95.

These subtle differences in the WW95 and T95 oxygen yields are interesting 
in their own right (\eg Langer 1997), and 
are certainly encapsulated by the 30\% error budget adopted by T97.
Unfortunately, a more fundamental flaw made by T97 was in assuming that \it
all \rm of the uncertainty in the
oxygen yields could be accounted for by this 30\% factor.  The problem with
assuming that the T95 oxygen yields form a lower envelope, with WW95
forming the upper, can best be appreciated by referring to the two other primary
sources of SNe models, besides those of T95 and WW95 -- \ie
Langer \& Henkel (1995, hereafter LH95) and Arnett (1996, hereafter A96).

In Figure \ref{fig:fig1}, 
we can see that both LH95 and A96 agree (roughly) with one another,
\footnote{The treatment of convection, the adopted
$^{12}$C$(\alpha,\gamma)^{16}$O reaction rate, and the initial evolutionary
state of the models, were all significantly different in T95, than in either
LH95 or A96.  LH95 also considered sub-solar metallicity models and
self-consistent mass-loss, both of which the others did not.}
\it but \rm lie a factor of $\sim 3$ below the T95 and WW95 predictions.  This
factor of three uncertainty in the oxygen yields is a far cry from the 30\%
uncertainty advocated by T97, but is entirely in keeping with that found by
Langer (1997).  This latter reference is an invaluable resource for those
interested in properly assessing the origin of the large uncertainties in
different modelers' oxygen yields.  In the same vein, a useful comparison of
the primary differences in the model
input physics can be found in Table 2 of Gibson,
Loewenstein \& Mushotzky (1997).  At this point, we are \it not \rm advocating
one grid of oxygen yields over 
another, simply drawing attention to the fact that T97 have
underestimated its uncertainty.

Figure \ref{fig:fig1} reflects the present-day state-of-the-art, as far as
oxygen yields goes, and can be considered to be the modern equivalent of Wang \&
Silk's (1993) Figure 1.  The similarity of the Arnett (1978), Woosley \& Weaver
(1986), and Thielemann, Nomoto \& Hashimoto
(1994) oxygen yields, reflected in the latter
figure, is not in dispute, but it should also be readily apparent (from our
Figure \ref{fig:fig1}) that
supplementing the current analogs of these older models (\ie A96, WW95, and
T95, respectively) 
with the new models of LH95 demonstrates that the agreement is no better
than a factor of $\sim 2\rightarrow 3$ at a given initial mass, in agreement
with that found by Langer (1997).

\begin{figure}[ht]
\epsscale{1.0}
\plotone{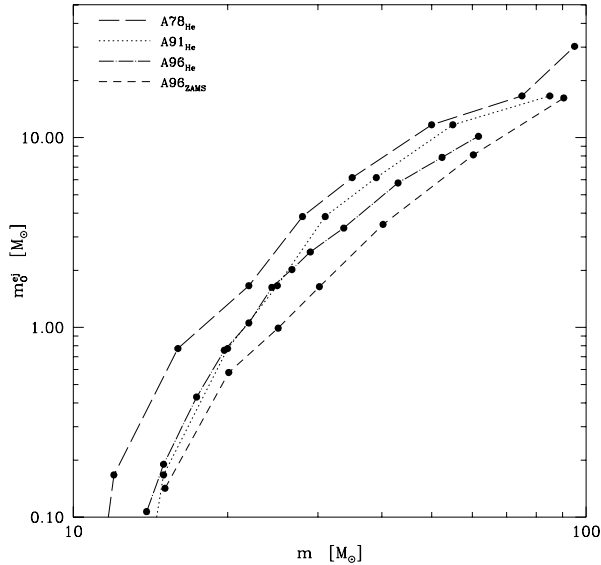}
\caption[fig2.eps]{
Oxygen yields as a function of progenitor mass, from the four compilations of
Arnett (1978, 1991, 1996).  Only the A96$_{\rm ZAMS}$ grid was evolved
self-consistently from the zero age main sequence (ZAMS), 
while the remainder were
evolved helium cores with an assumed \it a posteriori \rm ZAMS mass-He core
mass relation.
\label{fig:fig2}}
\end{figure}

Figure \ref{fig:fig2} parallels that of Figure \ref{fig:fig1}, but restricts
itself to the oxygen yields from the
published compilations of Arnett (1978,1991,1996), hereafter A78, A91, and A96,
respectively.  The subscript `He' or `ZAMS' denotes the initial evolutionary
state of the stellar models - pure Helium core or zero age main sequence.  The
former implies that a ZAMS mass-He core mass relation has been applied \it a
posteriori\rm, similar to the models of T95, whereas the latter avoid such
complications by modeling the evolution self-consistently.  We include a
parallel analysis using each of the Arnett generations as (i) A78 has already
played a large part in the analysis of Wang \& Silk (1991), and for
completeness sake should therefore be included here; (ii) A78 and A91
are identical in all respects, save the adopted He core mass-ZAMS mass
relation, a useful comparison in and of itself; (iii) A96$_{\rm He}$ and
A96$_{\rm ZAMS}$ are similar in all respects, save initial evolutionary
state, thereby allowing us to better appreciate this crucial difference in the
modeling.  Langer (1997) has previously alluded to the danger of comparing
yields from models evolved from the ZAMS versus those evolved from He cores; we
are now in a position to quantify this ``danger'', as applied to the
determination of the IMF upper mass limit.

A further assumption inherent to the T97 analysis is that the iron yields
are invariant and best represented by those of T95.  Iron
predictions are particularly problematic as the exact location of the mass-cut
leads to enormous uncertainties (Thielemann \etal 1996).  For the sake of
self-consistency in the input physics though, we have not made the same \it a
priori \rm assumption regarding iron's invariance, and have attempted
to match the oxygen predictions with iron from the same grid.  This is easy in
the case of T95 and WW95, as they follow the evolution right through
core-collapse and explosive nucleosynthesis.  Of the different Arnett grids
described herein, only A91 included a specific entry for iron, so that is used
for each of A78, A91, and A96, hereafter.
Proper self-consistency with Arnett's
yields will be restored after he applies core collapse and explosive 
nucleosynthesis to the new A96 model grid.

\begin{deluxetable}{crrrr}
\footnotesize
\tablecaption{IMF-Weighted Yields\tablenotemark{a}
\label{tbl:tbl1}}
\tablewidth{4in}
\tablehead{
\colhead{Source\tablenotemark{b}} & \colhead{[Fe/H]\tablenotemark{c}} & 
\colhead{$<{\rm O}>$\tablenotemark{d}} & 
\colhead{$<{\rm Fe}>$\tablenotemark{d}} & 
\colhead{$<{\rm [O/Fe]}>$}
}
\startdata
T95$_{\rm He}$ & +0.0$\;\;$  & 1.810$\;\;$ & 0.091$\;\;$ & +0.42$\quad\;\;$ \nl
M92$_{\rm ZAMS}$ & -1.3$\;\;$  & 1.508$\;\;$ & n/a$\;\;\;$ & n/a$\quad\;\;\;$ \nl
WW95$_{\rm ZAMS}$& -4.0$\;\;$  & 1.445$\;\;$ & 0.087$\;\;$ & +0.34$\quad\;\;$ \nl
LH95$_{\rm ZAMS}$ & -1.0$\;\;$  & 0.841$\;\;$ & n/a$\;\;\;$ & n/a$\quad\;\;\;$ \nl
A78$_{\rm He}^{\rm O}$+A91$_{\rm He}^{\rm Fe}$    & +0.0$\;\;$  & 1.711$\;\;$ & 0.068\tablenotemark{e}$\;\;$ & +0.52$\quad\;\;$ \nl
A91$_{\rm He}^{\rm O}$+A91$_{\rm He}^{\rm Fe}$    & +0.0$\;\;$  & 1.189$\;\;$ & 0.068\tablenotemark{e}$\;\;$ & +0.37$\quad\;\;$ \nl
A96$_{\rm He}^{\rm O}$+A91$_{\rm He}^{\rm Fe}$    & +0.0$\;\;$  & 0.983$\;\;$ & 0.068\tablenotemark{e}$\;\;$ & +0.28$\quad\;\;$ \nl
A96$_{\rm ZAMS}^{\rm O}$+A91$_{\rm He}^{\rm Fe}$    & +0.0$\;\;$  & 0.678$\;\;$ & 0.068\tablenotemark{e}$\;\;$ & +0.12$\quad\;\;$ \nl
\enddata
\tablenotetext{a}{
IMF slope $x=1.35$, over the range $10\rightarrow 50$ M$_\odot$.}
\tablenotetext{b}{
Source of oxygen yields.}
\tablenotetext{c}{
Metallicity of models from which oxygen yields were derived.}
\tablenotetext{d}{
IMF-weighted yield mass, in M$_\odot$.
}
\tablenotetext{e}{
Arnett (1991) [Fe/H]=+0.0 iron yields.}
\end{deluxetable}

We are now in a position to anticipate some
of the conclusions of Section \ref{analysis2}.  To do so, we weight the
mass-dependent oxygen and iron yields of Figures \ref{fig:fig1} and
\ref{fig:fig2} by some
canonical IMF, which we will take to be of the slope favored by Salpeter (1955)
- \ie $x=1.35$ - with Type II SNe progenitors assumed to span the mass range
$10\rightarrow 50$ M$_\odot$.  The resulting mean IMF-weighted masses of oxygen
and iron (and their logarithmic ratios relative to the solar ratio) - \ie
$<{\rm O}>$, $<{\rm Fe}>$, and $<{\rm [O/Fe]}>$ - for each of the yield
``pairs'' discussed thus far, are listed in Table \ref{tbl:tbl1}.  The
subscripts ZAMS and He refer to the initial evolutionary state of the models -
\ie zero age main sequence or simple helium cores.

The factor of $\sim 3$ uncertainty due to
oxygen (and to a lesser extent, that due to iron) is immediately apparent in
columns 3 and 4, their combination leading to a factor of $\sim 2$
uncertainty in the IMF-weighted [O/Fe].

Recalling that T97, using the T95 yields,
assumed a metal-poor normalization for equation
\ref{eq:imf1} of [O/Fe]$_{\rm h}\approx +0.4$, and subsequently
found $m_{\rm U}\approx 50$ M$_\odot$, it should not be surprising to see that
our entry in Table \ref{tbl:tbl1}
for T95, generated using $m_{\rm U}=50$ M$_\odot$,
has an IMF-weighted [O/Fe] similar to T97's adopted halo
value.  Conversely, the A96 entries lie substantially below this
halo value.  Because [O/Fe] increases with increasing progenitor mass (see
Figure \ref{fig:fig1}), it should be apparent that to
recover an [O/Fe]$\approx +0.4$,
with the same IMF slope, will require a greater $m_{\rm U}$.  

\subsection{Quantifying $m_{\rm U}$'s Dependence Upon Yield Source}
\label{analysis2}

\begin{figure}[ht]
\epsscale{1.0}
\plotone{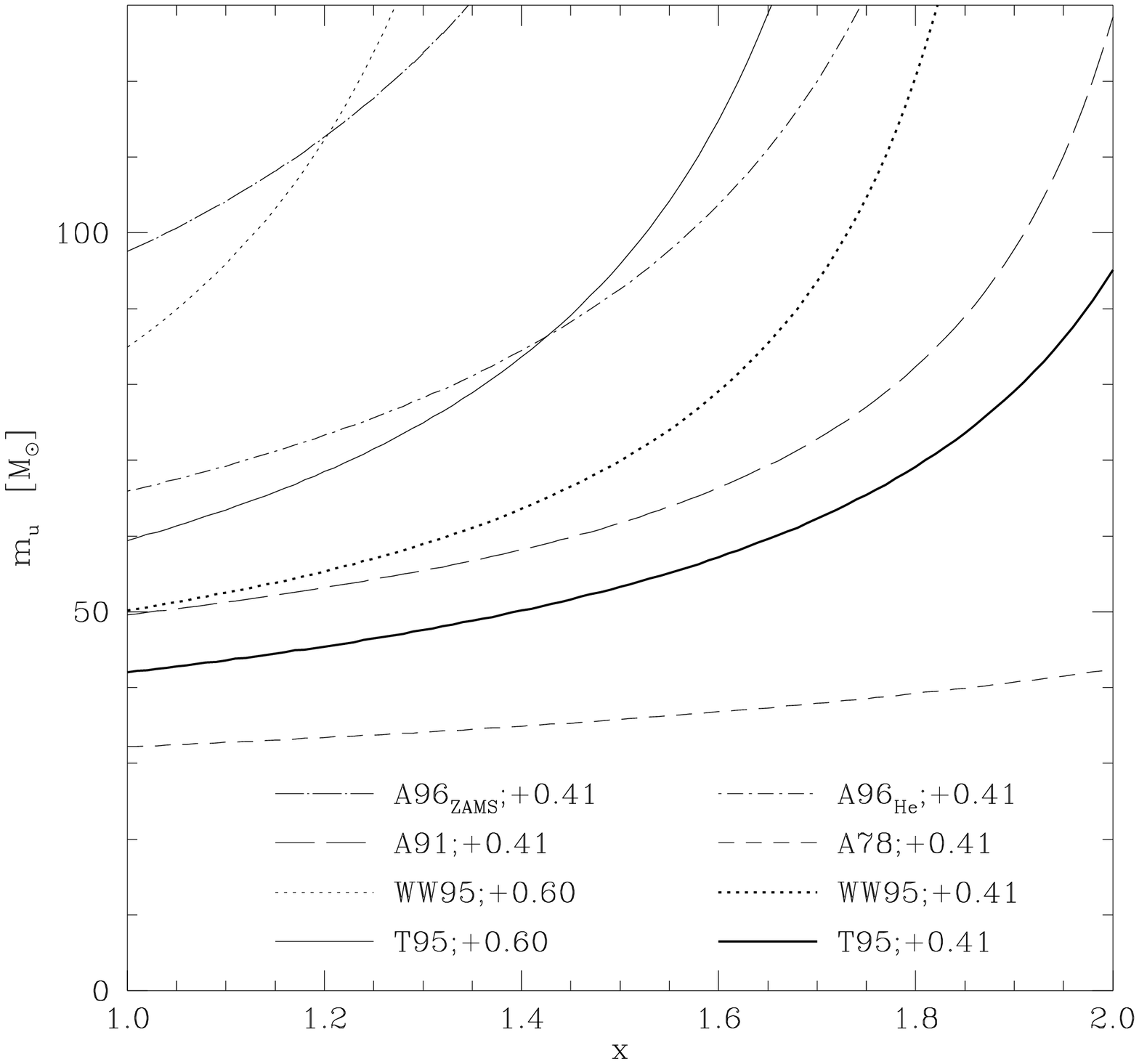}
\caption[fig3.eps]{
Upper mass limit $m_{\rm U}$,
as a function of IMF slope $x$, for the Type II SNe yield compilations of
Figures 1 and 2.
\label{fig:fig3}}
\end{figure}

For a given halo ``plateau'' normalization [O/Fe]$_{\rm h}$, the various yield
compilations of Figures \ref{fig:fig1} and \ref{fig:fig2}
(and Section \ref{analysis1a}) can now be
used in conjunction with equation \ref{eq:imf1} to determine how the upper mass
limit $m_{\rm U}$ varies as a function of IMF slope $x$.  The results, for 
a variety of models, are shown in Figure \ref{fig:fig3}.  Let us restrict
ourselves, for the time being, to the six curves which were generated assuming
the default halo normalization of [O/Fe]=+0.41.

The first model to note from Figure \ref{fig:fig3} is that given by the solid
curve, which effectively recovers the results of T97\footnote{The footnote to
Section \ref{analysis1a}
clarifies why there remains a residual $\sim 10$\% offset 
between the two.} -- \ie $m_{\rm U}\approx 48\rightarrow 62$ 
M$_\odot$, for reasonable
IMF power-law slopes spanning $x\approx 1.3$ (Salpeter 1955) to $x\approx 1.7$
(Kroupa \etal 1993).  Again, recalling our comments at the end of Section
\ref{analysis1a}, this low predicted value for $m_{\rm U}$ was
already anticipated, based upon the T95 yields possessing the greatest
IMF-weighted $<$[O/Fe]$>$, for the ``standard'' IMF employed in constructing
Table \ref{tbl:tbl1}.

The most important result to take from Figure \ref{fig:fig3}, though, is not so
much the behavior of the T95 curve \it alone\rm, but the parallel behavior 
for the models employing the alternate Type II yields -- \ie those denoted 
WW95 and A96,\footnote{We have not shown the results for the LH95 oxygen yields
in Figure \ref{fig:fig3}, as they parallel closely those of A96$_{\rm ZAMS}$.}
in the legend to the figure.  For a Salpeter (1955) IMF
slope (\ie $x=1.35$), the predicted upper mass limit for the T95 yields was 
$m_{\rm U}\approx 49$ M$_\odot$.  For comparison, the corresponding $m_{\rm U}$
for the WW95, A78 and A96$_{\rm ZAMS}$ yields 
is 61, 35, and 131 M$_\odot$, respectively, with the A91 and A96$_{\rm He}$
values lying between A78 and A96$_{\rm ZAMS}$.
Because the dependence of $m_{\rm U}$ upon IMF slope $x$ is stronger in the
non-T95 yields (especially for A96$_{\rm ZAMS}$), for the Kroupa \etal (1993) 
slope (\ie $x=1.7$) $m_{\rm U}$ is more strongly affected.  The
value favored by the T95 compilation (\ie
62 M$_\odot$) is replaced, in this case, by values
ranging from $\sim 95$ M$_\odot$, for WW95,\footnote{Because the maximum mass
considered by WW95 was 40 $M_\odot$, we make the (perhaps mistaken) assumption
that linear extrapolation beyond the grid extreme holds.
For example, such an assumption leads to a predicted oxygen
yield of 14 M$_\odot$ for an $m=100$ M$_\odot$ star.  If we were to
arbitrarily double this yield to $\sim$30 M$_\odot$, and set the iron yield
to a constant 0.075 M$_\odot$ for $m\ge 35$ M$_\odot$, we would find an 
approximate 5\%/15\% reduction in the predicted $m_{\rm U}$, for IMF slopes
$x=1.3/1.6$.  Even more extreme, if we increase the same oxygen yield
to 50 M$_\odot$, and set the iron yield to zero for $m\ge 35$ M$_\odot$, we
would find an approximate 15\%/30\% reduction in the predicted $m_{\rm U}$, for
IMF slopes $x=1.3/1.6$.  Obviously, self-consistent yield predictions for the
WW95 grid, for masses $m>40$ M$_\odot$, are required, although the previous
examples do provide a feel for maximum effect involved.}
to $\simgt 200$ M$_\odot$, for A96$_{\rm ZAMS}$.

To summarize, \it only if one assumes that the T95 yields are the definitive
representation of low-metallicity massive star nucleosynthetic yields can one
conclude that the upper mass limit to the solar neighborhood IMF is $m_{\rm
U}=40\rightarrow 60$ M$_\odot$.  \rm By 
adopting, in turn, the three other primary
yield sources (\ie WW95, LH95, and A96), we sample a much fairer representative
input physics ``parameter space'', leading to the more conservative
result that
we are able to constrain the \it lower \rm 
bound to the upper mass limit to be
$\sim 40$ M$_\odot$, but that the \it upper \rm bound remains effectively
unconstrained.
If we force the Salpeter (1955) IMF slope (\ie $x=1.35$) to hold, the
uncertainty in the yields still only allows us to constrain $m_{\rm U}$ to the
range $\sim 40\rightarrow 140$ M$_\odot$.  \it This yield-dependence of $m_{\rm
U}$'s determination was not appreciated in the original analysis by T97.\rm

It is interesting to question how one might recover a result of $m_{\rm
U}\approx 40\rightarrow 60$ M$_\odot$, when adopting the A96$_{\rm ZAMS}$ 
yields.  Assuming the oxygen as inviolate, it is apparent from inspection of
Table \ref{tbl:tbl1} that we are required to reduce the IMF-weighted iron yield
by a factor of $\sim 2$.  This cannot be of the form of a blanket,
mass-independent reduction, since we are constrained by the observations of SNe
1987a and 1993j (Thielemann \etal 1996).
Only by setting the iron yield to zero for $m\le 12$ M$_\odot$ and $m\ge 30$
M$_\odot$ (\ie retaining the 15, 20, and 25 M$_\odot$ iron predictions of A91,
but ignoring the contribution from other masses), can we reduce the A96$_{\rm
ZAMS}$ prediction for $m_{\rm U}$ from $\sim 130\rightarrow 200$ M$_\odot$ 
to that found using T95 (\ie $m_{\rm U}\approx 40\rightarrow 60$ M$_\odot$).

In passing, we note that linear interpolation for both oxygen and iron
were assumed in equation \ref{eq:imf1}.  If one adopts a
logarithmic interpolation, as Yoshii \etal (1996) did, the predicted values 
of $m_{\rm U}$ would be reduced by $\sim 10\rightarrow 25$\%, from those shown
in Figure \ref{fig:fig3}.

\subsection{Quantifying $m_{\rm U}$'s Dependence Upon the Halo Normalization}
\label{analysis3}

T97 did not consider how their conclusions might be influenced by the adopted
mean ``halo'' [O/Fe]$_{\rm h}$ (equation \ref{eq:imf1}).  Recall that the solid
curve in Figure \ref{fig:fig3} supported $m_{\rm U}\approx 48\rightarrow 62$
M$_\odot$, for IMF slopes $x\approx 1.3\rightarrow 1.7$, for the T95
yields and a normalization [O/Fe]$_{\rm h}=+0.41$.

If instead of adopting T97's favored halo normalization, we took that from
Bessell \etal (1991) - \ie [O/Fe]$_{\rm h}=+0.60$ - we would recover the
``T95;+0.60'' curve of Figure \ref{fig:fig3}.
In other words, this $\sim 0.2$ dex (\ie $\sim 60$\%)
higher normalization, for the T95 yields, increases the predicted $m_{\rm U}$
by $\sim 60$\% (from 48 M$_\odot$ to 75 M$_\odot$,
for $x=1.3$) to $\sim 140$\% (from 62 M$_\odot$ to 148 M$_\odot$, for $x=1.7$).
The same conclusion is reached when we adopt the WW95 yields, as shown by the
``WW95;+0.60'' curve of Figure \ref{fig:fig3}.
A general rule-of-thumb for Salpeter (1955) IMF slopes (\ie $x=1.35$)
is that \it 
a given percentage increase in the halo normalization, is accompanied
by the same percentage increase in the predicted $m_{\rm U}$, \rm regardless of
yield source.  For slopes $x\approx 1.7$, the increase in $m_{\rm U}$ with
increasing normalization is generally $\sim 2\rightarrow 3\times$ greater.

If we simply take the halo normalization to be [O/Fe]$_{\rm h}=+0.5\pm 0.1$, and
accept T97's allowable range of IMF slopes (\ie $x\approx
1.3\rightarrow 1.6$), Figure \ref{fig:fig3} would lead us to conclude that
$m_{\rm U}\approx 40\rightarrow 140$ M$_\odot$
better represents the valid range of $m_{\rm U}$,
\it for the T95 yields\rm -- we feel that this would have been a more 
realistic range than
the $m_{\rm U}\approx 40\rightarrow 60$ M$_\odot$ claimed by T97.

\section{Summary}
\label{summary}

T97 have recently revitalized interest in using IMF-weighted
Type II SNe yields as a direct probe of said IMF's upper mass limit 
$m_{\rm U}$, by 
comparison with the observed abundance ratios in metal-poor Galactic stars.
The beauty of this technique lies partially in its simplicity -- for a given
IMF slope, there is effectively only one free parameter -- the yield source.
Adopting the T95 yields, T97 found $m_{\rm U}=40\rightarrow 60$
M$_\odot$, for reasonable IMF slopes (\ie $x=1.3\rightarrow 1.6$).

The primary concern we have regarding T97's analysis lies in their
underlying assumption that \it all \rm of the uncertainty in the stellar model
physics could be encapsulated in a 30\% (0\%) error budget for oxygen (iron).
It should be obvious from Figures \ref{fig:fig1} and \ref{fig:fig2},
and Langer (1997), for example,
that this assumption is incorrect, and that a more realistic error budget would
allow for up to an order of magnitude greater leeway.  Stellar models are
simply not developed to the level that is inherently assumed by T97 --
convection, overshooting, mass-loss, reaction rates, metallicity,
C/O-core masses, fallback onto the remnant -- each conspire to increase the
uncertainties to the degree reflected by the yields shown in Figures
\ref{fig:fig1} and \ref{fig:fig2}.

A secondary concern is T97's inherent
assumption that the halo normalization [O/Fe]$_{\rm
h}=+0.41$ has no associated uncertainty.  Since values as high as [O/Fe]=+0.6
are still favored by some, this $\sim 60$\% uncertainty should be taken into
account.  For a Salpeter (1955) slope, there is (roughly) a one-to-one
correspondence between the halo normalization uncertainty and the corresponding
predicted upper mass limit uncertainty.

While we \it do \rm
agree with T97 that the \it lower \rm limit to $m_{\rm U}$ is $\sim
40$ M$_\odot$ (or $\sim 30$ M$_\odot$, if we adopt the extreme A78 yields), 
our more realistic exploration of input physics ``space''
demonstrates that we simply cannot constrain the \it upper \rm limit to 
any useful accuracy.  Taken
together, we can only conclude that, by this technique alone, $m_{\rm U}\simgt
40$ M$_\odot$, for IMF slopes $x=1.3\rightarrow 1.6$.
Fixing the IMF slope to that of Salpeter (1955), we can only constrain $m_{\rm
U}$ to lie somewhere between $\sim 40$ M$_\odot$ and $\sim 140$ M$_\odot$.
Finally, it would appear to be difficult to reconcile any $m_{\rm U}\simlt 100$
M$_\odot$ with either the A96$_{\rm ZAMS}$ or WW95;+0.60 halo normaliztion
grids.

While promising (provided
existing discrepancies in Type II SNe yields are eliminated), at the
present time, unfortunately, this technique, by itself,
does not substantially improve or constrain
our understanding of
the upper mass limit to the solar neighborhood IMF.
 
\acknowledgments

We wish to thank Takuji Tsujimoto and Sylvia Becker
for a number of helpful correspondences.
BKG acknowledges the financial support of an NSERC Postdoctoral Fellowship.


\end{document}